\documentclass[onecolumn,showpacs,preprintnumbers,elsart]{revtex4}
\usepackage{makeidx}
\usepackage{amssymb}
\usepackage{amsmath}
\usepackage{mathrsfs}
\usepackage{graphicx}
\usepackage{dcolumn}
\usepackage{bm}
\usepackage[center]{subfigure}
\usepackage{color}

\begin{document}

\title{Cross-symmetric dipolar-matter-wave solitons in double-well chains }
\author{Zhiwei Fan$^{1}$, Yuhan Shi$^{1}$, Yan Liu$^{1}$}
\email{lycalm@scau.edu.cn}
\author{ Wei Pang$^{2}$, Yongyao Li$^{1,3}$}
\email{yongyaoli@gmail.com}
\author{Boris A. Malomed$^{4,5,1}$}
\affiliation{$^{1}$Department of Applied Physics, South China Agricultural University,
Guangzhou 510642, China \\
$^{2}$ Department of Experiment Teaching, Guangdong University of
Technology, Guangzhou 510006, China.\\
$^{3}$School of Physics and Optoelectronic Engineering, Foshan University,
Foshan 528000, China\\
$^{4}$Department of Physical Electronics, School of Electrical Engineering,
Faculty of Engineering, Tel Aviv University, Tel Aviv 69978, Israel\\
$^{5}$Laboratory of Nonlinear-Optical Informatics, ITMO University, St.
Petersburg 197101, Russia}

\begin{abstract}
We consider a dipolar Bose-Einstein condensate trapped in an array of
two-well systems with an arbitrary orientation of the dipoles relative to
the system's axis. The system can be built as a chain of local traps sliced
into two parallel lattices by a repelling laser sheet. It is modelled by a
pair of coupled discrete Gross-Pitaevskii equations, with dipole-dipole
self- and cross-interactions. When the dipoles are not polarized
perpendicular or parallel to the lattice, the cross-interaction is
asymmetric, replacing the familiar symmetric two-component discrete solitons
by two new species of cross-symmetric ones, \textit{viz}., on-site- and
off-site-centered solitons, which are strongly affected by the orientation
of the dipoles and separation between the parallel lattices. A very narrow
region of intermediate asymmetric discrete solitons is found at the boundary
between the on- and off-site families. Two different types of solitons in
the $\mathcal{PT}$-symmetric version of the system are constructed too, and
stability areas are identified for them.
\end{abstract}

\maketitle

\section{Introduction}

Bose-Einstein condensates (BECs) composed of dipolar atoms and molecules is
a broad research area in low-temperature physics. This type of BEC,
dominated by anisotropic long-range magnetic or electric dipole-dipole
interactions (DDIs), is significantly different from usual condensates,
whose intrinsic dynamics is determined by ``point-blank" inter-atomic
collisions. Studies of dipolar BEC have produced a large number of specific
experimental and theoretical results, which have been summarized in Refs.
\cite{Griesmaier2007,Lahaye2009,Baranov}.

In addition to the fact that the atomic or molecular dipoles can be
polarized by external dc electric or magnetic fields, the sign of the DDI
can be switched by the ac rotating field \cite{Giovanazzi2002}. These
features lend dipolar BECs a great deal of tunability. In addition to the
use of atoms or molecules carrying permanent magnetic or electric moments,
condensates can be made of particles with dipole moments locally induced by
the same dc fields which polarize the moments, the latter setting with
spatially nonuniform fields being especially interesting \cite{YLi2016}.
These properties enhance the potential offered by dipolar BECs for
fundamental and applied studies. One significant direction in these studies
is the use of dipolar condensates for emulation of various phenomena which
occur in a more complex form in other physical media, such as rotons \cite%
{rotons}, ferrofluidity \cite{Saito2009,Richter2005}, Faraday waves \cite%
{Nath2010}, supersolids \cite{Buhler2011}, anisotropic superfluidity \cite%
{Ticknor2011}, anisotropic collapse \cite{collapse},
mesoscopic droplets stabilized by quantum fluctuations \cite{drops},
spin-orbit coupling in dipolar media \cite{SOC-dipolar}, and others \cite%
{Klawunn2009,Muller2011,Wilson2011,Gawryluk2011}.

Another noteworthy ramification is the use of collective nonlinear modes in
dipolar BEC for the creation of solitons in nonlocal media. This topic was
originally introduced in optics, where nonlocal nonlinearities of other
types occur \cite{nonlocal-reviews,Krolikowski2000}. Various forms of bright
\cite{bright}, dark \cite{dark}, vortex \cite{vortex} and discrete \cite%
{discrete} solitons were predicted in dipolar condensates. Recently, stable
two-dimensional solitons were predicted in the dipolar BEC with spin-orbit
coupling \cite{SOC}. In addition to BEC, it was demonstrated that the DDI
can create solitons in the ultracold bosonic gas of the Tonks-Gigardeau type
\cite{TG}.

A specific phenomenon which can be realized in dipolar BEC is the
spontaneous symmetry breaking (SSB), and the related phase transition, alias
the symmetry-breaking bifurcation, of modes created by such long-range
anisotropic interactions. The SSB is a ubiquitous effect, which occurs in
all areas of nonlinear physics \cite{SSB}. Because the nonlinearity often
creates solitons, a natural subject of the analysis is the SSB of solitons
in symmetric systems. In particular, many theoretical and experimental
results on this subject have been reported in photonic and matter-wave
settings, where the nonlinearity is an inherent feature, and symmetry is
frequently provided by dual-core or double-well structures \cite{SSB2}. The
SSB for solitons in systems with local nonlinearity has been studied in
detail theoretically, \cite{Trillo1988}-\cite{Li2011}, including discrete
systems, which represent parallel arrays of coupled waveguides \cite%
{Herring2007}. The study of the SSB for solitons in systems with nonlocal
interactions is a problem of considerable interest too, as the nonlocality
strongly affects the outcome of the competition between the nonlinear
self-focusing of the fields and linear mixing between them, which leads to
the SSB when the nonlinearity strength exceeds a critical value \cite{SSB2}.
Thus far, only few works have addressed this topic. In particular, the SSB
transformation of solitons in the dual-core coupler with nonlocal optical
nonlinearity of the thermal type was considered in Ref. \cite{Shi2012}.
Unlike the optical systems, dipolar BEC features not only the intra-core
nonlocal nonlinearity, but also the inter-core DDI, which makes the
situation essentially different, as was first shown in the model for the
effectively 1D dual-core coupler filled by the dipolar condensate \cite%
{YLi2013}. In that work, different types of SSB, sub- and supercritical ones
(i.e., symmetry-breaking phase transitions of the first and second kinds,

respectively) were demonstrated, as the result of the competition of the
DDIs with strong or weak hopping between the cores. However, the analysis
was performed in Ref. \cite{YLi2013} only for a single polarization of the
dipoles, namely, along the cores. However, the external magnetic field can
polarize the dipoles in any direction; once the dipoles are not parallel or
perpendicular to the core, nonlocal cross-interaction induced by the DDIs
becomes asymmetric, breaking the usual type of the solitons' symmetry.

The aim of the present work is to explore what kind of symmetry may be
featured by two-component discrete solitons for oblique orientation of the
dipoles. We demonstrate that, unless the dipole moments are oriented
strictly perpendicular or parallel to the system's axis, the solitons'
shapes become uneven (spatially asymmetric), which makes it necessary to
modify the definition of the symmetry between the soliton's components,
replacing it by \textit{cross-symmetry}. Actually, two types of
cross-symmetry are found for different sets of the system's parameters, see
Eqs. (\ref{newrelationship}) and (\ref{newrelationship2}) below. To the best
our knowledge, discrete solitons with such types of the symmetry were not
studied before.

The rest of the paper is structured as follows. The two-component discrete
model is introduced in Sec. II, and the cross-symmetry of solitons in it,
controlled by the orientation of the dipoles, is studied in Sec. III. In
Sec. IV, we introduce a further generalization of the system, by lending it
the $\mathcal{PT}$ symmetry (represented by spatially separated and mutually
balanced gain and loss). The paper is concluded by Sec. V.

\section{The model}

\begin{figure}[tbp]
\centering{\label{fig1a} \includegraphics[scale=0.25]{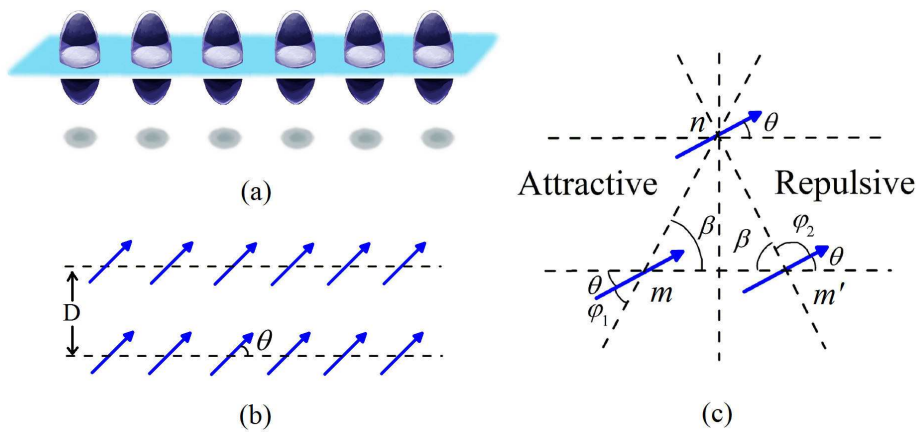}}
\caption{(Color online) (a) The system: a chain of tunnel-coupled traps for
BEC,\ sliced by a repelling laser sheet into a stack of double wells, i.e.,
a pair of parallel lattices. (b) The orientation of dipoles in the system. $%
D $ is the distance between the parallel lattices. (c) A typical example of
the setting for the asymmetric nonlocal cross-interaction. In this case, the
$m\leftrightarrow n$ and $m^{\prime }\leftrightarrow n$ DDI is attractive
and repulsive, respectively, at $\protect\varphi _{1}<54.7^{\circ }$ and $%
\protect\varphi _{2}>54.7^{\circ }$. }
\label{sketch}
\end{figure}

We consider a chain of two-well systems, into which the dipolar BEC is
loaded, as schematically shown in Fig. \ref{sketch}(a). It can be built as
the usual quasi-one-dimensional (1D) lattice \cite{Markus}, split into a
pair of parallel ones by an additional repulsive (blue-detuned) laser sheet.
We consider configurations with different angles $\theta $ of the
orientation of the dipoles with respect to the lattice, as shown in Fig. \ref%
{sketch}(b).

In the tight-binding approximation \cite{tight,Panos}, the mean-field
dynamics of the condensate in this system is governed by the two-component
discrete Gross-Pitaevskii equation for wave functions $\tilde{\psi}_{n}$ and
$\tilde{\phi}_{n}$ of particles trapped in local potential wells:
\begin{eqnarray}
&&i{\frac{d}{dt}}\tilde{\psi}_{n}=-{\frac{C}{2}}(\tilde{\psi}_{n+1}+\tilde{%
\psi}_{n-1})+\left[ \sigma |\tilde{\psi}_{n}|^{2}+\sum_{m\neq n}\left(
F_{nm}|\tilde{\psi}_{m}|^{2}+G_{nm}|\tilde{\phi}_{m}|^{2}\right) \right]
\tilde{\psi}_{n}-J\tilde{\phi}_{n},  \notag \\
&&i{\frac{d}{dt}}\tilde{\phi}_{n}=-{\frac{C}{2}}(\tilde{\phi}_{n+1}+\tilde{%
\phi}_{n-1})+\left[ \sigma |\tilde{\phi}_{n}|^{2}+\sum_{m\neq n}\left(
F_{nm}|\tilde{\phi}_{m}|^{2}+G_{mn}|\tilde{\psi}_{m}|^{2}\right) \right]
\tilde{\phi}_{n}-J\tilde{\psi}_{n}.  \label{baseEq}
\end{eqnarray}%
Here, $C$ and $J$ are, respectively, the coupling constants (determined by
the respective hopping rates) along the lattice and between the parallel
ones, $\sigma $ is the strength of the contact nonlinearity, while $F_{nm}$
and $G_{nm}$ are DDI\ kernels, which account, severally, for the nonlocal
self- and cross-interactions in the coupled Gross-Pitaevskii equations:
\begin{eqnarray}
&&F_{nm}=%
\begin{cases}
0, & (m=n) \\
{(1-3\cos ^{2}\theta )/|m-n|^{3}} & (m\neq n)%
\end{cases}%
,  \label{F} \\
&&G_{nm}=%
\begin{cases}
(1-3\sin ^{2}\theta )/D^{3} & (m=n) \\
{\left[ 1-3\cos ^{2}\varphi _{1}\right] /[D^{2}+(m-n)^{2}]^{3/2}} & (m<n) \\
{\left[ 1-3\cos ^{2}\varphi _{2}\right] /[D^{2}+(m-n)^{2}]^{3/2}} & (m>n)%
\end{cases}%
,  \label{G}
\end{eqnarray}%
where $D$ is the scaled separation between the parallel lattices, $\varphi
_{1}=\beta -\theta $, and $\varphi _{2}=\pi -(\beta +\theta )$ [see in Fig. %
\ref{sketch}(c)], with $\beta \equiv \arccos \left( |m-n|/\sqrt{%
D^{2}+(m-n)^{2}}\right) $. Recently, a similar configuration was considered
as an Ising model with long-range interactions, which does not include the
transverse hopping, i.e., with $C=0$ \cite{we2017}.

Stationary states are looked for in the usual form,
\begin{equation}
(\tilde{\psi}_{n},\tilde{\phi}_{n})=(\psi _{n},\phi _{n})e^{-i\mu t},
\label{stationary}
\end{equation}%
where $(\psi _{n},\phi _{n})$ are stationary wave functions, and $\mu $ is a
real chemical potential. Two-component solitons are characterized by their
total norm,%
\begin{equation}
P=P_{\psi }+P_{\phi }\equiv \sum_{n=-N/2}^{n=N/2}\left( |\tilde{\psi}%
_{n}|^{2}+|\tilde{\phi}_{n}|^{2}\right) ,  \label{P}
\end{equation}%
which is a dynamical invariant of Eq. (\ref{baseEq}).

For $\theta =0$ or $\pi /2$, matrix $G_{nm}$ given by Eq. (\ref{G}) is
symmetric, with $G_{nm}=G_{mn}$. When $0<\theta <\pi /2$, this property is
broken, which makes nonlocal cross-interaction asymmetric, see Fig. \ref%
{sketch}(c). Obviously, at $\theta =0$ \textit{vertical }(alias
inter-lattice) interactions, i.e., the onsite DDI between condensate
droplets trapped in the two potential wells belonging to the parallel
lattices, is repulsive, while the \textit{horizontal} DDI along each lattice
is attractive. With the increase of $\theta $, the vertical\textit{\ }%
interaction vanishes at
\begin{equation}
\theta _{1}=\arcsin \left( 1/\sqrt{3}\right) \approx 0.196\pi \approx
35.3^{\circ },  \label{theta1}
\end{equation}%
while the horizontal DDI remains attractive. Another special angle,
\begin{equation}
\theta _{2}=\arccos \left( 1/\sqrt{3}\right) \approx 0.304\pi \approx
54.7^{\circ },  \label{theta2}
\end{equation}
corresponds to the vanishing of the horizontal interaction, while the
vertical DDI is attractive.

Obviously, at $\theta =0$ and $\pi /2$, symmetric discrete solitons obey the
spatial-symmetry condition,
\begin{equation}
\phi _{-n}=\phi _{n},\psi _{-n}=\psi _{n}.  \label{relationship}
\end{equation}%
However, when $\theta $ is different from $0$ and $\pi /2$, shapes of the
two components are not spatially even because, as mentioned above, $G_{nm}$
is not a symmetric matrix anymore. In the following section we introduce
another type of symmetry which two-component discrete solitons may feature
in this case. To the best of our knowledge, the asymmetric
cross-interactions corresponding to asymmetric $G_{nm}$ were not considered
previously.

\section{Cross-symmetric solitons}

\begin{figure}[tbp]
\centering{\label{fig2a1} \includegraphics[scale=0.3]{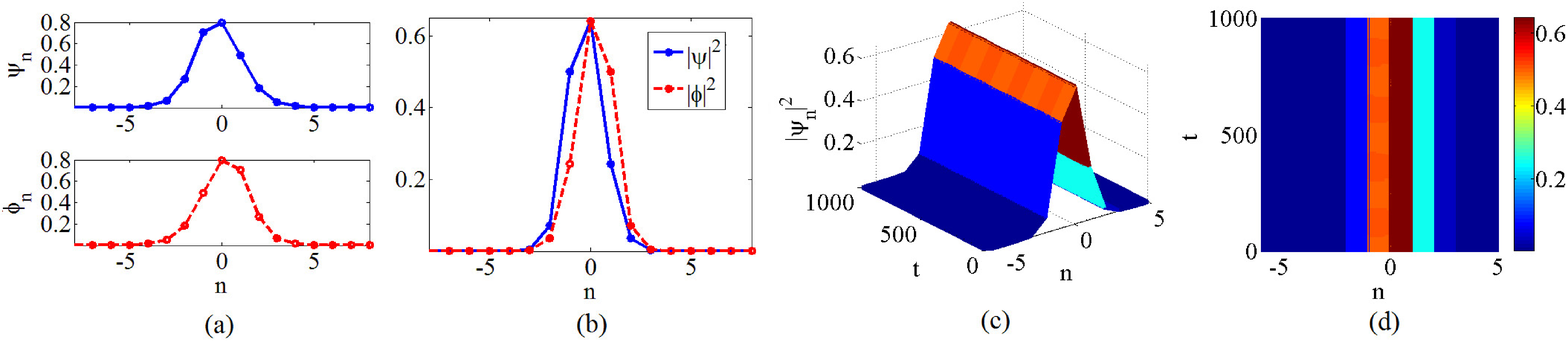}}
\caption{(Color online) A typical example of a cross-symmetric soliton for $%
(P,D,\protect\theta )=(3,0.4,0.196\protect\pi )$. (a) Shapes of two
components of the solitons. (b) Juxtaposition of their density profiles. The
blue and red curves are mirror images of each other. (c) Simulations of
weakly-perturbed evolution of the cross-symmetric soliton, which makes its
stability evident. (d) A top view of the configuration displayed in (c).}
\label{on-site}
\end{figure}

\begin{figure}[tbp]
\centering{\label{fig3a} \includegraphics[scale=0.35]{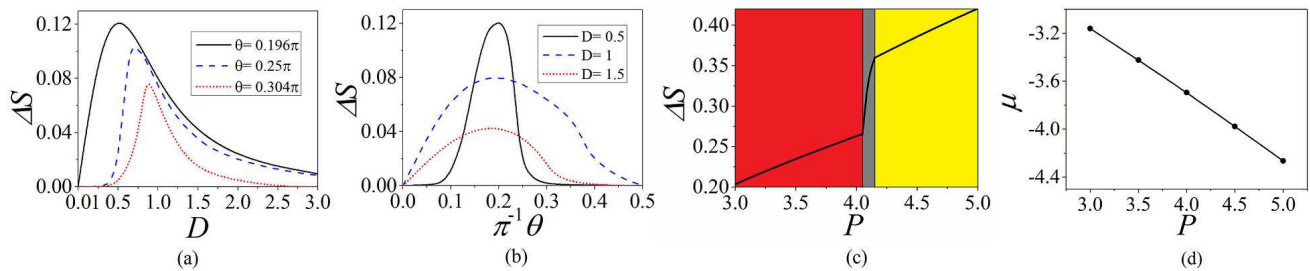}}
\caption{(a,b) The cross-symmetry$\ $measure $\Delta {S}$, defined as per
Eq. (\protect\ref{DeltaS}), versus $D$ and $\protect\theta $ for discrete
solitons featuring the cross-symmetry of the on-site type, at a fixed total
power, $P=1.9$. (c) $\Delta S$ versus $P$ (black solid curve) when $D=0.5$
and $\protect\theta =0.196\protect\pi $. Here, the line $\Delta S(P)$
traverses areas populated by three types of the discrete solitons: on-site
cross-symmetric (red), intermediate state (gray), and off-site
cross-symmetric (the yellow area). (d) Dependence $\protect\mu (P)$ for $%
D=0.5$ and $\protect\theta =0.196\protect\pi $, which clearly satisfies the
VK\ (Vakhitov-Kolokolov) criterion, $d\protect\mu /dP<0$. }
\label{delta}
\end{figure}

\begin{figure}[tbp]
\centering{\label{fig2a2} \includegraphics[scale=0.3]{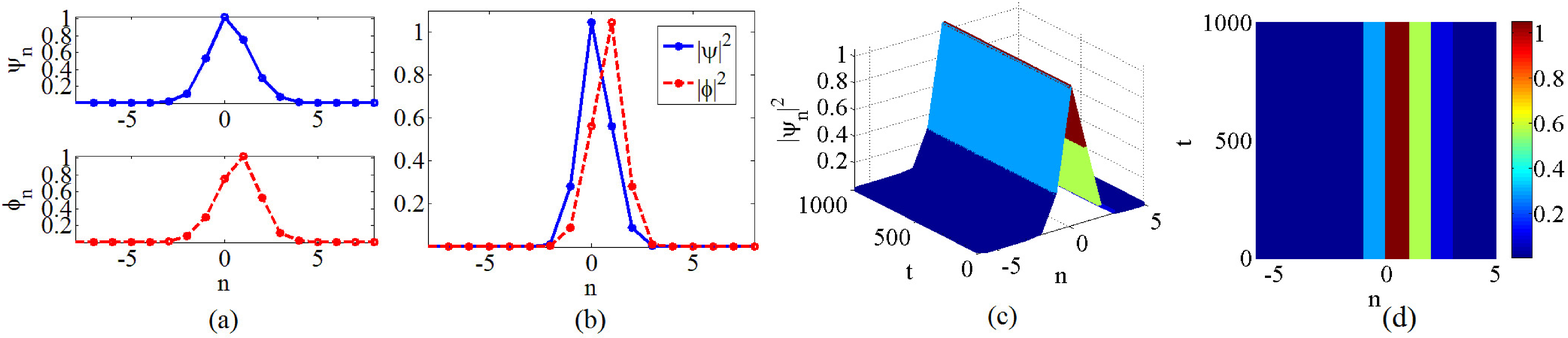}}
\caption{(Color online) A typical example of discrete solitons featuring the
cross-symmetry of the off-site type, see Eq. (\protect\ref{newrelationship2}%
), for $(P,D,\protect\theta )=(4,0.4,0.196\protect\pi )$. Panels have the
same meaning as in Fig. \protect\ref{on-site}.}
\label{off-site}
\end{figure}

To focus on effects induced by the DDIs, we drop the contact nonlinearity in
the system, by setting $\sigma =0$ in Eq. (\ref{baseEq}), and assume equal
horizontal and vertical hopping rates, scaling both to be unity: $C=J=1$.
The remaining control parameters are $P$, $D$ and $\theta $, i.e., the total
norm of the solitons [see Eq. (\ref{P})], separation between the lattices,
and the orientation of the dipoles, respectively. Here, we consider the case
of $0<\theta <\pi /2$, in which $G_{nm}$ is not symmetric, as said above.

The only feasible approach to the study of the present system may be based
on numerical methods. Discrete solitons in models with nearest-neighbor
interactions can be explored by means of a variational approximation \cite%
{VA}; however, it cannot be developed in an analytically tractable form for
lattices with long-range interactions.

Figure \ref{on-site} displays a typical example of a two-component discrete
soliton, with $(P,D,\theta )=(3,0.4,0.196\pi )$, obtained numerically by
means of the imaginary-time method \cite{Chiofalo,Jianke1,Jianke2}. The
figure corroborates that the two components of the soliton indeed do not
obey symmetry conditions (\ref{relationship}); nevertheless, they satisfy
the definition of the \emph{cross-symmetry}:
\begin{equation}
\phi _{n}=\psi _{-n},\psi _{n}=\phi _{-n},  \label{newrelationship}
\end{equation}%
which is compatible with Eq. (\ref{baseEq}) in the case of asymmetric
cross-interaction matrix $G_{nm}$. Note that locations of maxima of both
components coincide in Fig. \ref{on-site}.

It is relevant to stress that all the soliton families considered below,
which may be characterized by the respective dependences $\mu(P)$, satisfy
the well-known Vakhitov-Kolokolov (VK) criterion \cite{VVKK}, $d\mu /dP<0$
[a typical example of dependence $\mu (P)$ is displayed in Fig. \ref{delta}%
(d)], which is a necessary condition for stability of solitons against small
perturbations. While this criterion was originally established for solitons
in continuum media \cite{VVKK}, its generalization for two-component
discrete solitons is known too \cite{VanGorder}. In fact, the results
reported below demonstrate that the VK criterion is sufficient for the
stability of discrete solitons in the present system (except for its $%
\mathcal{PT}$-symmetric extension, which is introduced in Section IV). In
this connection, it is relevant to mention that, although configurations
with dipoles oriented perpendicular to the system's direction tends to be
the most stable one \cite{Giamarchi}, we here demonstrate that the discrete
solitons with oblique orientations are stable too.

It follows from Eq. (\ref{newrelationship}) that cross-symmetric solitons
have equal norms of the their components, $P_{\psi }=P_{\phi }$. The
cross-symmetry is quantified by the on-site mismatch between the components,
defined as
\begin{equation}
\Delta S={\frac{1}{P}}\sum_{n=-N/2}^{n=N/2}{\LARGE |}\left( |{\psi }%
_{n}|^{2}-|{\phi }_{n}|^{2}\right) {\LARGE |}.  \label{DeltaS}
\end{equation}%
For usual symmetric solitons Eq. (\ref{DeltaS}) yields $\Delta S=0$, as it
follows from Eq. (\ref{relationship}). A larger magnitude of $\Delta S$
corresponds to a stronger mismatch between the two components.

The cross-symmetry may suffer \textit{spontaneous breaking}, similar to the
above-mentioned SSB\ phenomena. This effect will be signaled by emergence of
soliton solutions with $P_{\psi }\neq P_{\phi }$, and will be considered
elsewhere.

To dependence of the cross-symmetry degree (\ref{DeltaS}) on the
inter-lattice separation $D$ and orientation $\theta $ is displayed in Fig. %
\ref{delta}, which shows that $\Delta S$ attains a maximum at finite values
of $D$ and $\theta $. According to the figure, the cross-symmetry is well
pronounced around the maximum, in parameter intervals $0.35<D<0.55$ and $%
0.18\pi <\theta <0.22\pi $. Further consideration of solitons in this area
with the increase of the total norm reveals another variety of the
cross-symmetry, different from that defined by Eq. (\ref{newrelationship}).
A typical example of the new variety is displayed in Fig. \ref{off-site}.
Comparing it with the counterpart displayed above in Fig. \ref{on-site}, we
find that maxima of the two components are separated in Fig. \ref{off-site}
by one lattice site, with the respectively modified cross-symmetry defined
as
\begin{equation}
\phi _{n}=\psi _{1-n},\psi _{n}=\phi _{1-n},  \label{newrelationship2}
\end{equation}%
cf. Eq. (\ref{newrelationship}). It is possible to say that cross-symmetry
axes, corresponding to definitions (\ref{newrelationship}) and (\ref%
{newrelationship2}), are set, severally, on-site and off-site (at the
midpoint between two sites in the latter case), therefore we refer to these
two varieties as on- and off-site cross-symmetries, respectively.

\begin{figure}[tbp]
\centering{\label{fig4a} \includegraphics[scale=0.35]{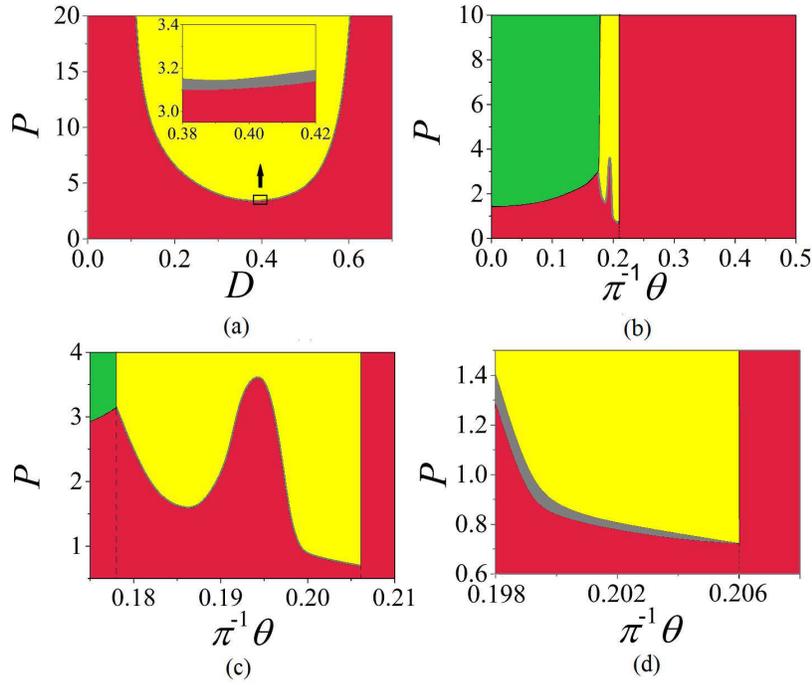}}
\caption{(Color online) Existence regions of stable discrete cross-symmetric
discrete solitons of the on- and off-site are shown in parameter planes by
red and yellow colors, respectively. In panel (a), orientation $\protect%
\theta =0.196\protect\pi $ is fixed, and the off-site (yellow) area exists
between $D=0.1$ and $0.62$. The inset in (a) displays a zoom around $D=0.4$,
with the minuscule gray area representing the intermediate state. In panel
(b), $D=0.4$ is fixed. In the green region, asymmetric solitons are found.
The off-site solitons exist in the region between $\protect\theta =0.177%
\protect\pi $ and $0.206\protect\pi $, which is displayed at a larger scale
in panel (c). Panel (d) is a zoom of a very small gray area, where the
intermediate solitons are found at $0.198\protect\pi <\protect\theta <0.206%
\protect\pi $.}
\label{typeofcross}
\end{figure}

\begin{figure}[tbp]
\centering{\label{fig2a3} \includegraphics[scale=0.3]{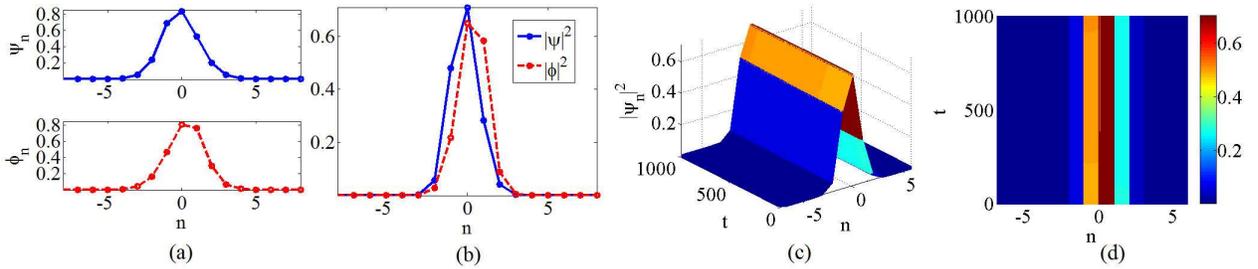}}
\caption{(Color online) A typical example of a stable asymmetric discrete
soliton, intermediate between cross-symmetric ones of the on- and off-site
types, for $(P,D,\protect\theta )=(3.15,0.4,0.196\protect\pi )$. Panels have
the same meaning as in Figs. \protect\ref{on-site} and \protect\ref{off-site}%
.}
\label{mix-state}
\end{figure}

Existence areas of the cross-symmetric discrete solitons of these two types
in the $\left( P,D\right) $ and $\left( P,\theta \right) $ planes are
displayed in Fig. \ref{typeofcross}, which shows that the off-site
cross-symmetric solitons exist in finite areas [their boundaries are
vertical in panels \ref{typeofcross}(b-d) up to the accuracy of the
numerical results]. Along the border between these areas, there is a very
narrow band, shown as a gray strip, which is filled by discrete solitons of
an intermediate type. They do not feature any explicit symmetry, except for
the equality between the total norms of the two components, $P_{\psi
}=P_{\phi }$, see a typical example in Fig. \ref{mix-state}. The asymmetric
intermediate states account for a continuous transition between the
cross-symmetric discrete solitons of the on- and off-site types, being as
stable as their cross-symmetric counterparts are. The continuity of the
transition is made evident in Fig. \ref{delta}(c) by the dependence of the
cross-symmetry degree, $\Delta S$ [defined by Eq. (\ref{DeltaS})], on the
total norm, going across areas occupied by these three types of the discrete
solitons.

\begin{figure}[tbp]
\centering{\label{interactionfor} \includegraphics[scale=1]{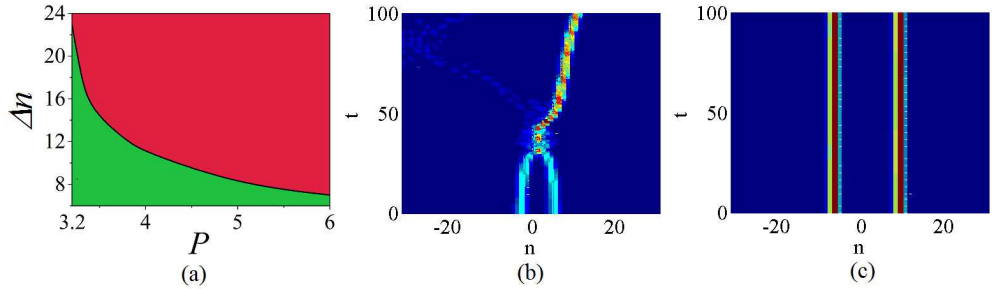}}
\caption{(Color online) (a) In the green area of the plane of $\left(
P,\Delta n\right) $, two in-phase solitons with identical powers $P$,
separated by initial distance $\Delta n$, merge into a single excited mode
[as shown in (b) for $P=4,\Delta n=8$], while in the red area the solitons
stay put [see panel (c), for $P=4,\Delta n=16$]. Other parameters are $D=0.4$%
, $\protect\theta =0.196\protect\pi $. }
\label{interactsoliton}
\end{figure}

We have also studied interaction between two cross-symmetric solitons,
originally separated by distance $\Delta n$. If $\Delta n$ is smaller than a
certain critical value, $\left( \Delta n\right) _{\mathrm{cr}}$, which
corresponds to the boundary between green and read areas in Fig. \ref%
{interactsoliton}(a), the solitons with zero phase shift between them
attract each other and merge into a single excited (oscillating) state, as
shown in In Fig. \ref{interactsoliton}(b). At $\Delta n>\left( \Delta
n\right) _{\mathrm{cr}}$, the pinning force from the underlying lattice is
stronger than the attraction, and the solitons stay in the initial positions
[see Fig. \ref{interactsoliton}(c)]. It is easy to understand why $\left(
\Delta n\right) _{\mathrm{cr}}$ strongly grows with the decrease of $P$, as
seen in Fig. Fig. \ref{interactsoliton}(a): for small $P$, the broad
solitons are quasi-continuum modes, for which the force of pinning to the
lattice is exponentially small \cite{Alik}.

\section{Two-component discrete solitons in the system with cross-$\mathcal{%
PT}$-symmetry}

Theoretical studies of many linear- and nonlinear-wave systems may be
naturally extended by adding the $\mathcal{PT}$-symmetry, i.e., spatially
symmetric distributions of globally balanced gain and loss terms \cite%
{Bender,PT}. In particular, much interest has been drawn to solitons in $%
\mathcal{PT}$-symmetric systems \cite{PT-review}. In term of matter waves,
the gain and loss represent symmetrically placed and mutually balanced
sources and sinks of coherent atoms. Although the experimental realization
of sources in BEC may not be easy, the theoretical analysis of BEC-$\mathcal{%
PT}$ systems has attracted considerable interest, see, e.g., Refs. \cite%
{Wunner,EPL,PT-RMP,Abdel}. In particular, a model applying the $\mathcal{PT}$
symmetry to dipolar BEC trapped in a symmetric double potential well was
proposed recently \cite{Wunner-DD}. In this section, we aim to develop the $%
\mathcal{PT}$-symmetric version of the system based on Eq. (\ref{baseEq}),
which, in particular, will provide, as far as it is known to us, the first
example of a discrete $\mathcal{PT}$-symmetric system with long-range
interactions.

Thus, Eq. (\ref{baseEq}) is replaced by
\begin{eqnarray}
&&i{\frac{d}{dt}}\tilde{\psi}_{n}=-{\frac{C}{2}}(\tilde{\psi}_{n+1}+\tilde{%
\psi}_{n-1})+\left[ \sigma |\tilde{\psi}_{n}|^{2}+\sum_{m\neq n}\left(
F_{nm}|\tilde{\psi}_{m}|^{2}+G_{nm}|\tilde{\phi}_{m}|^{2}\right) \right]
\tilde{\psi}_{n}-J\tilde{\phi}_{n}+i\kappa \tilde{\psi}_{n},  \notag \\
&&i{\frac{d}{dt}}\tilde{\phi}_{n}=-{\frac{C}{2}}(\tilde{\phi}_{n+1}+\tilde{%
\phi}_{n-1})+\left[ \sigma |\tilde{\phi}_{n}|^{2}+\sum_{m\neq n}\left(
F_{nm}|\tilde{\phi}_{m}|^{2}+G_{mn}|\tilde{\psi}_{m}|^{2}\right) \right]
\tilde{\phi}_{n}-J\tilde{\psi}_{n}-i\kappa \tilde{\phi}_{n}.
\label{PT-baseEq}
\end{eqnarray}%
where $\kappa >0$ is the coefficient accounting for the gain and loss of
atoms in the first and second components, respectively. The continuum limit
of Eq. (\ref{PT-baseEq}), with dominant local nonlinearity, resembles known
models of nonlinear $\mathcal{PT}$-symmetric couplers, in which the $%
\mathcal{P}$ transformation amounts to swapping the two coupled cores, one
carrying gain and the other being lossy \cite{Driben,Lisbon,Alexeeva}. In
the general case of $0<\theta <\pi /2$ considered in this work, which
corresponds to the asymmetric interaction matrix $G_{nm}$, Eq. (\ref%
{PT-baseEq}) realizes the \textit{cross-}$\mathcal{PT}$ symmetry, in the
sense of the cross symmetry defined as Eq. (\ref{PTrelationship2}) below.
Because the latter definition is actually tantamount to the $\mathcal{P}$
transformation along discrete coordinate $n$, Eq. (\ref{PT-baseEq})
effectively defines a 2D $\mathcal{PT}$-symmetric system (cf. the definition
of $\mathcal{CPT}$ symmetry proposed in Ref. \cite{EPL}).

In the continuous model of the $\mathcal{PT}$-symmetric coupler with cubic
intra-core nonlinearity, stationary symmetric solutions take the general
form of \cite{Driben,Alexeeva}
\begin{eqnarray}
&&\tilde{\psi}(x)=e^{-i\mu t}f(x)\exp \left[ {(i/2)}\arcsin \kappa \right] ,
\notag \\
&&\tilde{\phi}(x)=e^{-i\mu t}f(x)\exp \left[ -{(i/2)}\arcsin \kappa \right] ,
\label{PTcon}
\end{eqnarray}%
where $\mu $ is a real chemical potential of the solutions, and real $f(x)$
is a solution of the single continuous equation without the $\mathcal{PT}$
terms, for the same $\mu $. Accordingly, a $\mathcal{PT}$-symmetric solution
of the discrete system is looked for as
\begin{eqnarray}
&&\tilde{\psi}_{n}=e^{-i\mu t}f_{n}\exp \left[ {(i/2)}\arcsin \kappa \right]
\equiv e^{-i\mu t}\psi _{n},  \notag \\
&&\tilde{\phi}_{n}=e^{-i\mu t}f_{n}\exp \left[ -{(i/2)}\arcsin \kappa \right]
\equiv e^{-i\mu t}\phi _{n}.  \label{PTdis}
\end{eqnarray}

Equation (\ref{PTdis}) suggests that $\mathcal{PT}$-symmetric states exist
when $\kappa <1$, satisfying the following relations:
\begin{eqnarray}
&&\mathrm{Re}[\psi _{n}]=\mathrm{Re}[\psi _{-n}]=\mathrm{Re}[\phi _{n}]=%
\mathrm{Re}[\phi _{-n}],  \notag \\
&&\mathrm{Im}[\phi _{n}]=\mathrm{Im}[\phi _{n}]=-\mathrm{Im}[\phi _{n}]=-%
\mathrm{Im}[\phi _{-n}],  \notag \\
&&|\psi _{n}|^{2}=|\psi _{-n}|^{2}=|\phi _{n}|^{2}=|\phi _{-n}|^{2},
\label{PTrelationship}
\end{eqnarray}%
cf. Eq. (\ref{relationship}). In our system, Eq. (\ref{PTrelationship})
holds when the nonlocal cross-interaction is symmetric, i.e., with a
symmetric matrix $G_{nm}$, which is correct for the orientation angles $%
\theta =0$ or $\pi /2$. As said above, in the case of asymmetric $G_{nm}$,
i.e., for $0<\theta <\pi /2$, the spatial symmetry is replaced by the
on-site or off-site cross-symmetry, i.e., discrete cross\textit{-}$\mathcal{%
PT}$ symmetric solution should be subject to constraints
\begin{eqnarray}
&&\mathrm{Re}[\psi _{n}]=\mathrm{Re}[\phi _{-n}],  \notag \\
&&\mathrm{Im}[\phi _{n}]=-\mathrm{Im}[\phi _{-n}],  \notag \\
&&|\psi _{n}|^{2}=|\phi _{-n}|^{2}.  \label{PTrelationship2}
\end{eqnarray}%
or
\begin{eqnarray}
&&\mathrm{Re}[\psi _{n}]=\mathrm{Re}[\phi _{1-n}],  \notag \\
&&\mathrm{Im}[\phi _{n}]=-\mathrm{Im}[\phi _{1-n}],  \notag \\
&&|\psi _{n}|^{2}=|\phi _{1-n}|^{2}.  \label{PTrelationship3}
\end{eqnarray}

\begin{figure}[tbp]
\centering{\label{fig6a1} \includegraphics[scale=0.3]{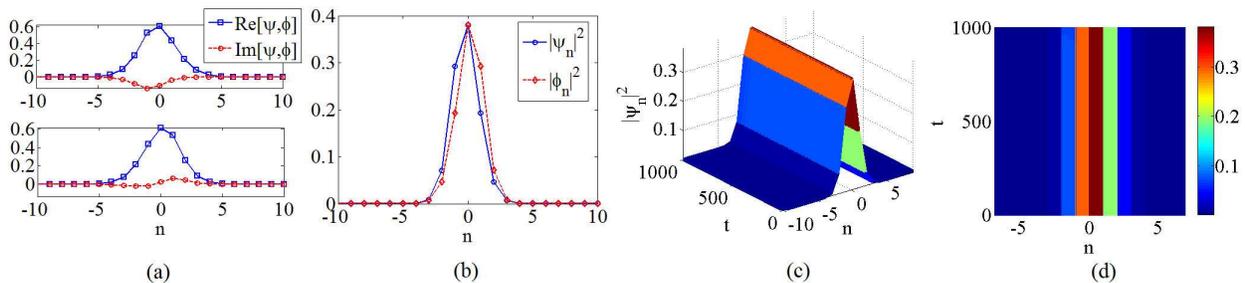}}
\caption{(Color online) A typical example of stable cross-$\mathcal{PT}$%
-symmetric solitons of the on-site type, for $(P,D,\protect\theta ,\protect%
\kappa )=(2,0.4,0.196\protect\pi ,0.2)$. Panel (a) displays the real and
imaginary parts of both field components. Panels (b),(c) and (d) have the
same meaning as in Figs. \protect\ref{on-site}, \protect\ref{off-site} and
\protect\ref{mix-state}.}
\label{PTon}
\end{figure}

\begin{figure}[tbp]
\centering{\label{fig6a2} \includegraphics[scale=0.3]{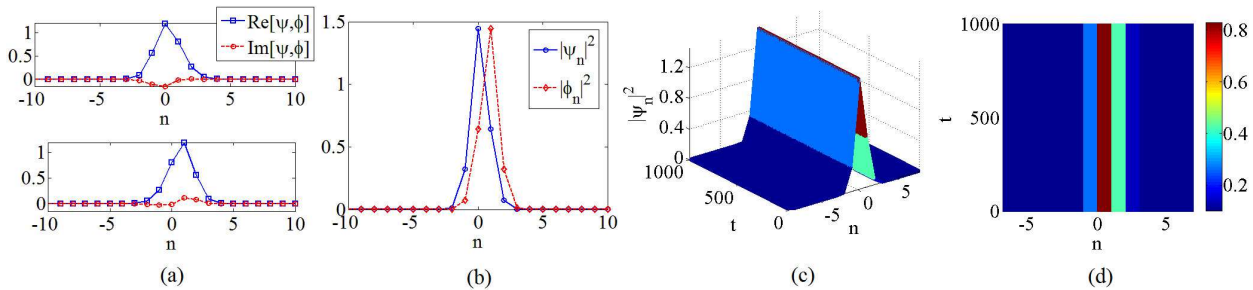}}
\caption{(Color online) A typical example of stable cross-$\mathcal{PT}$%
-symmetric solitons of the off-site type, for $(P,D,\protect\theta ,\protect%
\kappa )=(5,0.4,0.196\protect\pi ,0.1)$. Panels have the same meaning as in
Fig. \protect\ref{PTon}. }
\label{PToff}
\end{figure}

\begin{figure}[tbp]
\centering{\label{fig5a} \includegraphics[scale=0.2]{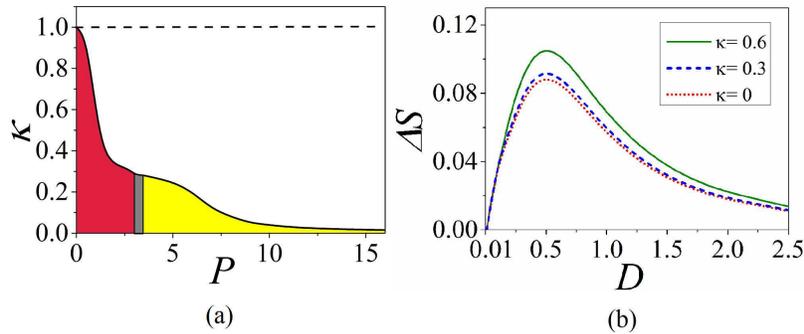}}
\caption{(Color online) (a) The stability map for discrete solitons in the
cross\textit{-}$\mathcal{PT}$-symmetric system for $(D,\protect\theta %
)=(0.4,0.196\protect\pi )$. The red and yellow areas represents stability
regions for the cross\textit{-}$\mathcal{PT}$-symmetric solitons of the on-
and off-site types, respectively. The intermediate gray stripe, $3<P<3.3$,
is populated by unstable asymmetric solitons. All solutions are unstable in
the white area. (b) The cross-symmetry measure (\protect\ref{DeltaS}) vs. $D$%
, for the solitons of the on-site cross-$\mathcal{PT}$-symmetric type with $%
(P,\protect\theta )=(1.5,0.196\protect\pi )$, cf. Fig. \protect\ref{delta}%
(a) for the conservative system. }
\label{PTcurve}
\end{figure}

\begin{figure}[tbp]
\centering{\label{fig6a3} \includegraphics[scale=0.3]{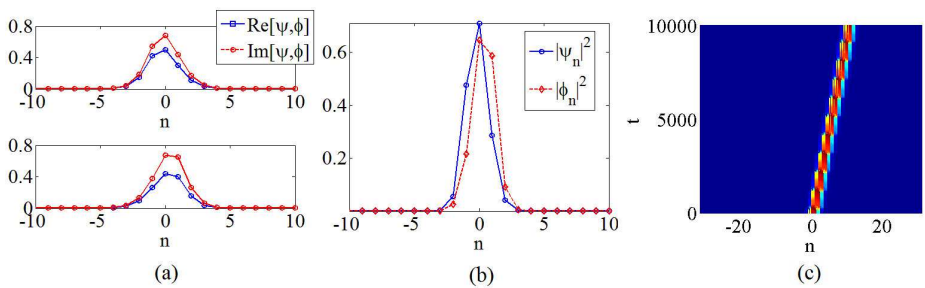}}
\caption{(Color online) A typical example of an unstable asymmetric soliton
found in the cross-$\mathcal{PT}$-symmetric system for $(P,D,\protect\theta ,%
\protect\kappa )=(3.15,0.4,0.196\protect\pi ,0.05)$. Panels have the same
meaning as in Figs. \protect\ref{PTon} and \protect\ref{PToff}. Panel (c)
clearly demonstrates that the instability turns the stationary solitons into
an effectively stable moving breather.}
\label{PTasy}
\end{figure}

Figures \ref{PTon} and \ref{PToff} display, severally, typical examples of
cross-$\mathcal{PT}$-symmetric solitons of the of the on- and off-site
types. Similar to their counterpart in the conservative system, the off-site
cross-$\mathcal{PT}$-symmetric solitons exist, as stable states, only in a
narrow area in a vicinity of $\theta =0.196\pi $ and $D=0.4$. Further, Fig. %
\ref{PTcurve}(a) displays a stability area of these two types of solitons in
the $(\kappa ,P)$ plane for $\theta =0.196\pi $ and $D=0.4$. Sandwiched
between the two stability areas is a narrow (gray) stripe, where asymmetric
solitons of the intermediate type are found in the cross-$\mathcal{PT}$%
-symmetric system, see a typical example in Fig. \ref{PTasy}. Unlike their
counterparts in the conservative system, the asymmetric solitons are
unstable. However, on the contrary to unstable solitons in the usual $%
\mathcal{PT}$-symmetric systems \cite{PT-review}, they do not suffer a
blowup, as a result of the instability development. Instead, the instability
turns these soliton into robust moving breathers, as shown in Fig. \ref%
{PTasy}(c)

The influence of the strength of the gain-loss coefficient on the degree of
the cross-symmetry, $\Delta S$ [defined by the same expression (\ref{DeltaS}%
) as above] was studied too, as shown in Fig. \ref{PTcurve}(b), which
displays $\Delta S(D)$ dependences for different fixed values of $\kappa $,
at fixed $\theta =0.196\pi $ and $P=1.5$. It is seen that the increase of $%
\kappa $ slightly enhances the cross-$\mathcal{PT}$-symmetry, by making $%
\Delta S$ somewhat larger than in the case of $\kappa =0$.

Stable cross-$\mathcal{PT}$-symmetric solitons feature robust oscillations
under the action of random perturbations, see an example displayed in Fig.\ %
\ref{PTcurve}(a). As shown in Fig. \ref{PTcurve}(b), the peak frequency of
the power spectrum of the intrinsic oscilaltions of stable solitons of both
the on-site and off-site centered types increases with the growth of the
soliton's total power. Actually, it identifies the frequency of a dominant
\textit{internal mode} of the stable cross-$\mathcal{PT}$-symmetric solitons.

\begin{figure}[tbp]
\centering{\label{figadd2} \includegraphics[scale=1]{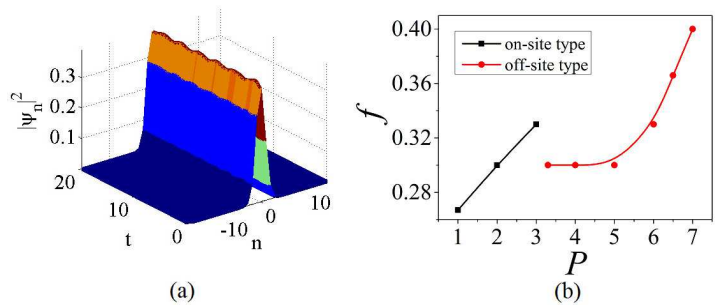}}
\caption{(Color online) (a) An example of robust oscillations of a stable
cross-$\mathcal{PT}$-symmetric soliton of the on-site-centered type, excited
by adding a white-noise perturbation to the initial conditions, with a $2\%$
relative amplitude. Parameters of this soliton are $(P,D,\protect\theta ,%
\protect\kappa )=(2,0.4,0.196\protect\pi ,0.1)$. (b) The frequency of the
intrinsic oscillations of the randomly-perturbed solitons of both on-site
and off-site-centered types, which corresponds to the maximum of the
respective power spectrum, versus the total power of the soliton. The other
parameters are $(D,\protect\theta ,\protect\kappa )=(0.4,0.196\protect\pi %
,0.1)$.}
\label{PToscillation}
\end{figure}

\section{Conclusion}

We have introduced the model of the chain of double-well potential traps for
dipolar BEC. In the tight-binding approximation, it amounts to a system of
two coupled discrete Gross-Pitaevskii equations with the long-range DDIs
(dipole-dipole interactions) determined by angle $\theta $ of the
orientation of the dipoles with respect to the system's axis. Except for the
limit cases of $\theta =0$ and $\theta =\pi /2$, the system, with the
spatially asymmetric DDI between the two parallel lattices, gives rise to
the cross-symmetric discrete two-component solitons of two different types,
on-site or off-site centered. These two families are stable, being separated
by a very narrow region populated by intermediate asymmetric discrete
solitons, which are stable too. Finally, we have extended the analysis by
adding the $\mathcal{PT}$-symmetry to the system. In this case, stability
areas for the cross-$\mathcal{PT}$-symmetric solitons of the on- and
off-site-centered solitons have been identified. The corresponding
intermediate asymmetric solitons are unstable, evolving into robust
breathers.

It may be interesting to extend the work to the consideration of
higher-order solitons, such as \textit{twisted} (spatially antisymmetric
\cite{Panos}) ones. A challenging possibility is to introduce a
two-dimensional version of the present discrete system.

\begin{acknowledgments}
We are indebted to Prof. L. Santos for valuable discussions. Z.W.F.
appreciates technical assistance provided by Ms. Jiaying Zou (Wuhan
Institute of Technology). This work was supported by the National Natural
Science Foundation of China through Grants 11575063, 61575041, and 11204089. The work
of B.A.M. is supported, in a part, by grant No. 2015616 from the joint
program in physics between Natural Science Foundation (US) and Binational
(US-Israel) Science Foundation. This author appreciates hospitality of the
Department of Applied Physics at the South China Agricultural University
(Guangzhou).
\end{acknowledgments}

\bibliographystyle{plain}
\bibliography{apssamp}

\end{document}